\documentclass[12pt]{article}
\usepackage{color,amssymb,amsthm,amsmath,fullpage,graphicx,hyperref,float,bm,array,booktabs}

\usepackage{fancyhdr}
\usepackage{xcolor}
\definecolor{dark-blue}{rgb}{0,0,0.45}
\hypersetup{colorlinks, linkcolor={dark-blue}, citecolor={blue}, urlcolor={blue}}

\numberwithin{equation}{section}

\def\Z{\mathbb{Z}}
\def\C{\mathbb{C}}
\def\H{\mathbb{H}}
\def\Qe{Q_8}
\def\E{\mathcal{E}}
\def\il{\mathbin{\between}}
\def\sq{\sqrt{-1}}
\DeclareMathOperator{\ord}{ord}
\DeclareMathOperator{\Aut}{Aut}

\definecolor{gold}{rgb}{0.85,0.66,0.0}

\newtheorem{theorem}{Theorem}[section]
\newtheorem{lemma}[theorem]{Lemma}
\newtheorem{conjecture}[theorem]{Conjecture}
\newtheorem{proposition}[theorem]{Proposition}
\newtheorem{corollary}[theorem]{Corollary}
\theoremstyle{definition}
\newtheorem{example}[theorem]{Example}
\newtheorem{definition}[theorem]{Definition}
\newtheorem{construction}[theorem]{Construction}

\title{A Recursive Construction of Perfect and Odd Perfect\\ Sequences of Unbounded Length over the Unit Quaternions}
\author{Sam Blake\\
\texttt{\small samuel [dot] thomas [dot] blake [at] gmail [dot] com}}
\date{14 November 2026}

\begin{document}

\maketitle

\begin{abstract}
We give a recursive construction, with closed-form index functions, of perfect and odd perfect
periodic autocorrelation sequences over the eight simple unit quaternions
$\Qe=\{\pm1,\pm i,\pm j,\pm k\}$ in every length $2^n$. The sequences are built, as in the
recursive perfect array construction of \cite{BlakeTower}, from \emph{twisted periodic extensions}:
a short sequence continued with period $M$ up to a right factor $u\in\Qe$. Over roots of unity the
twists cannot be regenerated by a one-dimensional recursion, consistently with Mow's conjecture;
over $\Qe$ we show that no single sequence of length $M\not\equiv0\pmod 4$ carries a
twist $\pm i,\pm j$ or $\pm k$ either, but that \emph{pairs} of sequences whose $i$- and
$(-i)$-twisted autocorrelations are complementary do. These $i$-pairs behave exactly like Golay
pairs: they double under interleaving with a right factor $j$ in place of a sign, every Golay pair of
length $n$ gives an $i$-pair of length $2n$, and they exist in length $6$ where no Golay pair exists.
An $i$-pair of length $M$ interleaves to an odd perfect sequence of length $4M$, and a perfect and an
odd perfect sequence of length $M$ interleave to a perfect sequence of length $4M$ with the array
orthogonality property for the divisor $2$. The $i$-pairs of length $2^t$ are $x^{(t)}_{r,a}=j^{w_t(r,a)}$,
where $w_t(r,a)$ counts the changes in the bit string formed by $r$ and the binary digits of $a$,
and the odd perfect sequence of length $2^{t+2}$ is $y^{(t)}_{2a+r}=j^{w_t(r,\,a\bmod 2^t)}\,
i^{(1-2r)\lfloor a/2^t\rfloor}$. The construction also gives odd perfect sequences over $\Qe$ in every length
$24\cdot2^s$ and perfect sequences in every length $6\cdot4^s$.
\end{abstract}

\section{Introduction}

Let $\omega=e^{2\pi\sq/n}$. A sequence $s=[s_0,s_1,\cdots,s_{M-1}]$ over the $n$th roots of unity
is \emph{perfect} if all of its off-peak periodic autocorrelations vanish. Mow's conjecture
\cite{Mow1993,Mow1996} asserts that perfect sequences over $n$ roots of unity of length greater than
$n^2$ do not exist, and the Heimiller--Frank sequences \cite{Heimiller1961,Frank1962} attain the
length $n^2$. Perfect sequences over the quaternions were introduced by Kuznetsov \cite{Kuz09} to
escape this barrier: the alphabet is fixed but non-commutative, and the quaternion analogues of the
Gaussian sums that drive the root-of-unity constructions are the sums $\sum_a i^{ca}$ of
\cite[Lemma~8.0.2]{Blake2017}. Kuznetsov \cite{Kuz10} gave a perfect sequence of length $5\,354\,228\,880$ over a
$24$-element alphabet of unit quaternions; the perfect sequences of lengths
$10,14,\dots,98$ over $\{\pm i,\pm j,\pm k\}$ of \cite[\S8.1]{Blake2017} were extended to unbounded
length by Barrera Acevedo and Hall through Lee sequences \cite{BAHall12}; and Barrera Acevedo and
Dietrich \cite{BADietrich18,BADietrich19} connected perfect sequences over the simple unit
quaternions $\Qe=\{\pm1,\pm i,\pm j,\pm k\}$ to Williamson sequences and to relative difference sets
in $C_n\times\Qe$. Chapter~8 of \cite{Blake2017} gave the index-function sequences
$s_a=i^{\lfloor a^2/2^{n-1}\rfloor}j^{\lfloor 2a^2/2^{n-1}\rfloor}$ (Construction~XXIII), which
are perfect over $\Qe$ for lengths $2^n\le64$ but not for $2^n\ge128$, and found a perfect
sequence of length $64$ over $\Qe$ with the array orthogonality property (AOP) for the divisor $8$,
the length being the square of the size of the alphabet as for the Heimiller--Frank sequences.
Bright, Kotsireas and Ganesh \cite{BKG20} constructed perfect and odd perfect sequences over $\Qe$
in every length $2^t$. Their construction passes through Williamson and nega Williamson sequences,
takes its odd perfect ingredient from Golay pairs, and gives no formula for the entries. Most
recently Bennett, Bright, Colinot and Nayak \cite{BBCN26} proved that quaternion-type and
Williamson-type sequences coincide, enumerated all perfect sequences over $\Qe\cup q\Qe$,
$q=\tfrac12(1+i+j+k)$, of length at most $21$, and related perfect quaternion sequences to the
circulant quaternionic Hadamard matrices arising in quantum communication.

In this paper we give a direct recursive construction over $\Qe$ which is closed under its own
operations and has explicit index functions. The organising idea is the one of
\cite{BlakeTower}, where a recursive construction of perfect $K^n\times K^n$ arrays over $K$ roots
of unity was obtained by working not with perfect arrays but with arrays of a \emph{type}
$v\in\Z_K^2$, that is arrays which are perfect only after a fractional linear phase, and whose
\emph{twisted periodic extensions} are the sub-arrays of the AOP decomposition of the next array in
the tower. In one dimension and over roots of unity that recursion fails at the first step, for want of
sequences of the type needed to lift, which is where Mow's barrier shows itself.
Over $\Qe$ the available twists are the eight elements of the group, and the picture is as follows.
No single sequence of length $M\not\equiv0\pmod4$ has a twist type $\pm i,\pm j$ or $\pm k$
(Proposition~\ref{notypeprop}), so the non-central twists cannot be carried by a sequence; but they
can be carried by a \emph{pair} of sequences with twists $i$ and $-i$ whose twisted autocorrelations
cancel, which we call an $i$-pair (Definition~\ref{pairdef}). The $i$-pairs are the quaternion
analogue of Golay pairs. They double under an interleaving in which the sign of Golay's construction
is replaced by a right factor $j$ (Theorem~\ref{doublethm}); every Golay pair of length $n$ gives an
$i$-pair of length $2n$ (Proposition~\ref{golayprop}); and they exist in length $6$, where no Golay
pair exists (Proposition~\ref{sixprop}). An $i$-pair of length $M$ interleaves to an odd perfect
sequence of length $4M$, and a perfect and an odd perfect sequence of length $M$ interleave to a
perfect sequence of length $4M$ with the AOP for the divisor $2$ (Theorem~\ref{liftthm}). The three
operations close, and Constructions~\ref{pairconstr}--\ref{perfconstr} of Section~\ref{towersec}
write out the resulting sequences of every length $2^n$. Section~\ref{rootsec} runs the same
construction over roots of unity, where the doubling is unavailable, and shows that the tower then
stops exactly at Mow's bound $K^2$. The lengths $24\cdot2^s$ (odd perfect)
and $6\cdot4^s$ (perfect) obtained from the length-$6$ seed appear to be new over $\Qe$.

All of the results below were found by computer before being proved; the program
\texttt{tower.py} accompanying the paper constructs every sequence mentioned and checks its
autocorrelations exactly. The three facts that rest on exhaustive search rather than proof are
stated as such where they occur.

\section{Definitions}

\subsection{Quaternions}
The real quaternions $\H=\{a+bi+cj+dk : a,b,c,d\in\mathbb{R}\}$ have $i^2=j^2=k^2=ijk=-1$, so that
$ij=k=-ji$, $jk=i=-kj$ and $ki=j=-ik$. The conjugate of $q=a+bi+cj+dk$ is $q^*=a-bi-cj-dk$,
$(pq)^*=q^*p^*$, and $q^*=|q|^2q^{-1}$. The \emph{simple unit quaternions}
$\Qe=\{\pm1,\pm i,\pm j,\pm k\}$ form the quaternion group, with centre $\{\pm1\}$; for $u\in\Qe$
we write $\ord u$ for its order, so $\ord1=1$, $\ord(-1)=2$ and $\ord u=4$ otherwise. Two elements of
$\Qe$ either commute or anticommute, and they commute if and only if one is central or both lie in
the same cyclic subgroup. The automorphism group $\Aut(\Qe)\cong S_4$ acts transitively on
$\{\pm i,\pm j,\pm k\}$ and fixes $\pm1$.

It is often convenient to write $q=\alpha+\beta j$ with $\alpha,\beta\in\C$; then $q^*=
\bar\alpha-\beta j$, $j\gamma=\bar\gamma j$ for $\gamma\in\C$, and $q\in\Qe$ exactly when one of
$\alpha,\beta$ is zero and the other lies in $\{\pm1,\pm i\}$. We call the entries of a sequence over
$\Qe$ lying in $\{\pm1,\pm i\}$ \emph{complex} and those in $\{\pm j,\pm k\}$ \emph{imaginary}, and
write $N_C(s)$, $N_J(s)$ for their numbers.

\subsection{Correlation}
A sequence $s=[s_0,s_1,\cdots,s_{M-1}]$ over $\Qe$ is indexed modulo $M$. As in
\cite[\S8]{Blake2017} and \cite{Kuz09,BKG20}, the periodic cross-correlation of two sequences $a,b$
of a common period $L$ for the shift $\tau$ is
\[
\theta_{a,b}(\tau)=\sum_{n=0}^{L-1}a_n\,b_{n+\tau}^*,
\]
with the conjugate on the second factor, and the periodic autocorrelation is
$\theta_s(\tau)=\theta_{s,s}(\tau)$; $\theta_s(\tau)$ with $\tau\ne0\bmod L$ is an off-peak
autocorrelation. (The other order, $\sum_n a_n^*b_{n+\tau}$, is the correlation of the entrywise
conjugate sequences; Kuznetsov \cite{Kuz09} proved that a quaternion sequence is right perfect if
and only if it is left perfect, see also \cite[\S2.3]{BBCN26}, so nothing depends on the choice.) Two sequences are \emph{orthogonal} if
$\theta_{a,b}(\tau)=0$ for all $\tau$. Since the entries are units, $\theta_s(0)=L$ and
\begin{equation}\label{herm}
\theta_s(-\tau)=\theta_s(\tau)^*.
\end{equation}
For a unit $v$ we have $\theta_{sv}=\theta_s$ and $\theta_{vs}(\tau)=v\,\theta_s(\tau)\,v^*$, so
multiplying all entries by a unit on the left or on the right preserves every property defined by the
vanishing of correlations.

A sequence $s$ of length $M$ is \emph{perfect} if $\theta_s(\tau)=0$ for all $\tau\ne0\bmod M$, and
\emph{odd perfect} \cite{Luke2003} if the nega-periodic extension $\hat s_{n+M}=-\hat s_n$, of period
$2M$, satisfies $\theta_{\hat s}(\tau)=0$ for all $\tau\ne0\bmod M$. For $\Qe$ the sums
$\sum_{a=0}^{4m-1}i^{ca}$, $c\ne0\bmod4$, of \cite[Lemma~8.0.2]{Blake2017} play the part of the
Gaussian sums; we shall use them in the form $\sum_{q=0}^{\ord w-1}w^q=0$ for $1\ne w\in\Qe$.

\subsection{Twisted extensions and types}
The following is the one-dimensional, quaternion form of the twisted extension and the type of
\cite{BlakeTower}.

\begin{definition}[twisted extension, type]\label{typedef}
Let $s$ be a sequence of length $M$ over $\Qe$ and $u\in\Qe$. The \emph{right $u$-twisted extension}
of $s$ is the sequence $\E_u[s]$ on $\Z$ with
\[
\E_u[s]_{qM+m}=s_m\,u^q,\qquad 0\le m<M,\ q\in\Z,
\]
of period $\ord(u)M$, so that $\E_u[s]_{n+M}=\E_u[s]_n\,u$ for all $n$. The \emph{$u$-twisted
autocorrelation} of $s$ is
\[
\vartheta^u_s(\tau)=\sum_{m=0}^{M-1}s_m\,\big(\E_u[s]_{m+\tau}\big)^*,\qquad\tau\in\Z,
\]
and $s$ has \emph{type $u$} if $\vartheta^u_s(\tau)=0$ for all $\tau\ne0\bmod M$.
\end{definition}

Type $1$ is perfection and type $-1$ is odd perfection, since $\E_{-1}[s]=\hat s$. Unwinding the
definition, $\vartheta^u_s(\tau)$ for $0<\tau<M$ is the periodic autocorrelation in which every term
that wraps around carries the factor $u^*$ between the two entries, $s_m\,u^*\,s_{m+\tau-M}^*$; for
$u=-1$ this is the nega-periodic autocorrelation. The basic relation between the twisted
autocorrelation of $s$ and the autocorrelation of its extension is the following.

\begin{lemma}\label{Tlem}
Let $s$ have length $M$, let $u\in\Qe$ and let $L$ be a multiple of $\ord(u)M$. Computing
$\theta_{\E_u[s]}$ over the period $L$,
\[
\theta_{\E_u[s]}(\tau)=\frac{L}{M}\,\vartheta^u_s(\tau)\qquad\text{for all }\tau\in\Z .
\]
In particular $s$ has type $u$ if and only if $\theta_{\E_u[s]}(\tau)=0$ for all
$\tau\ne0\bmod M$.
\end{lemma}
\begin{proof}
Write $n=qM+m$ with $0\le m<M$ and $0\le q<L/M$. Then $\E_u[s]_n=s_mu^q$ and
$\E_u[s]_{n+\tau}=\E_u[s]_{m+\tau}u^q$, so
\[
\E_u[s]_n\big(\E_u[s]_{n+\tau}\big)^*=s_m\,u^q(u^q)^*\big(\E_u[s]_{m+\tau}\big)^*
=s_m\big(\E_u[s]_{m+\tau}\big)^* ,
\]
and summing over $m$ and over the $L/M$ values of $q$ gives the claim.
\end{proof}

\begin{lemma}[symmetries]\label{symlem}
Let $s$ have length $M$ and $u\in\Qe$ with $u^2=-1$, and compute $\theta=\theta_{\E_u[s]}$ over the
period $4M$. Then for all $\tau$
\[
\theta(\tau+2M)=-\theta(\tau),\qquad\theta(2M-\tau)=-\theta(\tau)^* .
\]
Hence $\theta$ is determined by its values for $0\le\tau\le M$, and $s$ has type $u$ if and only if
$\vartheta^u_s(\tau)=0$ for $0<\tau<M$.
\end{lemma}
\begin{proof}
$\E_u[s]_{n+2M}=\E_u[s]_nu^2=-\E_u[s]_n$ gives the first identity, and with \eqref{herm}
$\theta(2M-\tau)=-\theta(-\tau)=-\theta(\tau)^*$.
\end{proof}

\subsection{Equivalences}
Left multiplication $s\mapsto vs$ by a unit preserves every type, as
$\vartheta^u_{vs}=v\,\vartheta^u_s\,v^*$. Right multiplication $s\mapsto sv$ sends type $u$ to
type $v^*uv$, because $\E_u[sv]_{qM+m}=s_mvu^q=s_m(vuv^*)^qv$, i.e.\ $\E_u[sv]=\E_{vuv^*}[s]\,v$. Reversal $s\mapsto[s_{M-1},\cdots,s_0]$ sends
type $u$ to type $u^*$. An automorphism $\phi\in\Aut(\Qe)$ applied entrywise satisfies
$\vartheta^{\phi(u)}_{\phi(s)}=\phi(\vartheta^u_s)$ and sends type $u$ to type $\phi(u)$. Hence the
six non-central types are equivalent to one another and we shall always take the twist to be $i$
when a non-central twist is needed; the types $\pm1$ are fixed by all of these operations.

\subsection{Polyphase components and the array orthogonality property}
Let $y$ be a sequence of length $2N$. Its \emph{polyphase components} are the sequences $c_0,c_1$ of
length $N$ with $y_{2n+r}=c_{r,n}$, and we write $y=c_0\il c_1$ for the interleaving. The array
associated with $y$ for the divisor $2$ in the sense of \cite[\S1.1]{Blake2017} is the $N\times2$
array whose columns are $c_0$ and $c_1$, and the AOP for the divisor $2$ requires the two columns to
be orthogonal and to form a periodic complementary pair. The computation that proves
\cite[Theorem~1.1.9]{Blake2017} is a reindexing of the autocorrelation sum and uses no
commutativity; for $d=2$ it reads
\begin{align}
\theta_{c_0\il c_1}(2\tau)&=\theta_{c_0}(\tau)+\theta_{c_1}(\tau),\label{ileven}\\
\theta_{c_0\il c_1}(2\tau+1)&=\theta_{c_0,c_1}(\tau)+\theta_{c_1,c_0}(\tau+1),\label{ilodd}
\end{align}
where the correlations on the right are computed over the period $N$. Interleaving commutes with
twisting: regarding $c_0\il c_1$ as a sequence of length $2N$ and $c_0,c_1$ as sequences of length
$N$, $\E_u[c_0\il c_1]=\E_u[c_0]\il\E_u[c_1]$.

\section{Orthogonality of commuting twists}\label{orthsec}

Over roots of unity, the twisted extensions of \cite{BlakeTower} have spectra supported on disjoint
cosets, which is why they are orthogonal under all shifts. The quaternion counterpart is the
following lemma, whose proof is a quaternion Gaussian sum.

\begin{lemma}[orthogonality]\label{orthlem}
Let $a,b$ be sequences of length $M$ over $\Qe$ and let $u,v\in\Qe$ commute with $u\ne v$. Let $L$
be a common multiple of $\ord(u)M$ and $\ord(v)M$. Then, over the period $L$,
\[
\theta_{\E_u[a],\E_v[b]}(\tau)=0\qquad\text{for all }\tau .
\]
\end{lemma}
\begin{proof}
With $n=qM+m$ as before, $\E_u[a]_n=a_mu^q$ and $\E_v[b]_{n+\tau}=\E_v[b]_{m+\tau}v^q$, so the
summand is
\[
a_m\,u^q(v^q)^*\big(\E_v[b]_{m+\tau}\big)^*=a_m\,(uv^*)^q\big(\E_v[b]_{m+\tau}\big)^* ,
\]
using $uv=vu$. The element $w=uv^*\in\Qe$ is not $1$ and its order divides $L/M$, so
$\sum_{q=0}^{L/M-1}w^q=0$ by \cite[Lemma~8.0.2]{Blake2017}. As $a_m$ and
$(\E_v[b]_{m+\tau})^*$ do not depend on $q$, the inner sum over $q$ vanishes for each $m$.
\end{proof}

Commutativity cannot be dropped: for $u=i$ and $v=j$ one has $\sum_{q=0}^3i^q(-j)^q=2-2k$. Twisted
extensions with anticommuting twists can and do have nonzero cross-correlations, and they lie
outside the calculus of this paper.

\section{Twisted pairs and the two lifts}\label{pairsec}

\begin{definition}[$(u,-u)$-pair]\label{pairdef}
Let $u\in\Qe$ and let $x_0,x_1$ be sequences of length $M$ over $\Qe$. We call $(x_0,x_1)$ a
\emph{$(u,-u)$-pair} if, computing over the common period $\ord(u)M$,
\[
\theta_{\E_u[x_0]}(\tau)+\theta_{\E_{-u}[x_1]}(\tau)=0\qquad\text{for all }\tau\ne0\bmod 2M .
\]
An $(i,-i)$-pair is called an \emph{$i$-pair}.
\end{definition}

By Lemma~\ref{Tlem} the condition says that the twisted autocorrelations $\vartheta^u_{x_0}$ and
$\vartheta^{-u}_{x_1}$ form a complementary pair. For $u=\pm1$ there is nothing new:

\begin{lemma}\label{pm1lem}
$(x_0,x_1)$ is a $(1,-1)$-pair if and only if $x_0$ is perfect and $x_1$ is odd perfect.
\end{lemma}
\begin{proof}
Over the period $2M$, $\theta_{\E_1[x_0]}(\tau+M)=\theta_{\E_1[x_0]}(\tau)$ and
$\theta_{\E_{-1}[x_1]}(\tau+M)=-\theta_{\E_{-1}[x_1]}(\tau)$. The conditions at $\tau$ and $\tau+M$,
$0<\tau<M$, therefore read $\theta_{\E_1[x_0]}(\tau)\pm\theta_{\E_{-1}[x_1]}(\tau)=0$ and force
both terms to vanish; at $\tau=M$ the sum is $2M-2M=0$ automatically.
\end{proof}

For $u=i$ the two terms need not vanish separately, and by Proposition~\ref{notypeprop} they
essentially never do. By Lemma~\ref{symlem} the sum
$S(\tau)=\theta_{\E_i[x_0]}(\tau)+\theta_{\E_{-i}[x_1]}(\tau)$ satisfies $S(\tau+2M)=-S(\tau)$
and $S(2M-\tau)=-S(\tau)^*$, so the $i$-pair condition is equivalent to
\begin{equation}\label{paircond}
\vartheta^i_{x_0}(\tau)+\vartheta^{-i}_{x_1}(\tau)=0\qquad\text{for }0<\tau\le M .
\end{equation}
The condition at $\tau=M$ has a combinatorial meaning. Since $\E_i[x]_{m+M}=x_mi$ we have
$\vartheta^i_x(M)=\sum_mx_m\,i^*\,x_m^*$, and $x_mi^*x_m^*$ is $-i$ when $x_m$ is complex and $+i$
when it is imaginary; so \eqref{paircond} at $\tau=M$ says
\begin{equation}\label{countcond}
N_C(x_0)-N_J(x_0)=N_C(x_1)-N_J(x_1).
\end{equation}

\begin{theorem}[the two lifts]\label{liftthm}
Let $x_0,x_1$ be sequences of length $M$ over $\Qe$.
\begin{enumerate}
\item[(i)] If $(x_0,x_1)$ is an $i$-pair, then $y=\E_i[x_0]\il\E_{-i}[x_1]$, regarded as a sequence
of length $4M$, is odd perfect.
\item[(ii)] If $x_0$ is perfect and $x_1$ is odd perfect, then $y=\E_1[x_0]\il\E_{-1}[x_1]$,
regarded as a sequence of length $4M$, is perfect, and has the AOP for the divisor $2$.
\end{enumerate}
\end{theorem}
\begin{proof}
Let $u=i$ in case (i) and $u=1$ in case (ii), and put $w=u^2\in\{\pm1\}$. The components
$c_0=\E_u[x_0]$ and $c_1=\E_{-u}[x_1]$ satisfy $c_{r,n+2M}=c_{r,n}(\pm u)^2=c_{r,n}w$, so the
interleaving on $\Z$ is $\E_w[y]$ with $y$ its first $4M$ terms, and we must show that
$\theta_{\E_w[y]}(\tau)=0$ for $\tau\ne0\bmod4M$. By \eqref{ileven},
$\theta_{\E_w[y]}(2\tau)=\theta_{c_0}(\tau)+\theta_{c_1}(\tau)$, which vanishes for
$\tau\ne0\bmod2M$ by Definition~\ref{pairdef} in case (i) and by Lemma~\ref{pm1lem} in case (ii). By
\eqref{ilodd}, $\theta_{\E_w[y]}(2\tau+1)$ is the sum of two cross-correlations of $c_0$ and $c_1$,
and both vanish by Lemma~\ref{orthlem}, since $u$ and $-u$ commute and differ. In case (ii) the two
columns of the array associated with $y$ for the divisor $2$ are $c_0$ and $c_1$ restricted to one
period $2M$; we have just shown that they are orthogonal and periodic complementary, which is the
AOP.
\end{proof}

Part (ii) is Theorem~3 of \cite{BKG20} with its symmetry hypotheses removed: no Williamson structure
is needed for perfection, only for the correspondence with Williamson matrices. Part (i) is the
step that \cite{BKG20} obtain from Golay pairs by a different route (Proposition~\ref{golayprop}
below); the point of stating it for arbitrary $i$-pairs is that the $i$-pairs are closed under the
doubling of Section~\ref{doublesec}, which is what makes the construction recursive. The same proof
shows that four sequences $x_0,x_1,x_2,x_3$ of length $M$ with
$\theta_{\E_1[x_0]}+\theta_{\E_i[x_1]}+\theta_{\E_{-1}[x_2]}+\theta_{\E_{-i}[x_3]}=0$ off the
multiples of $4M$ interleave by $4$ to a perfect sequence of length $16M$ with the AOP for the
divisor $4$, because the twists $1,i,-1,-i$ commute pairwise; we have not pursued this four-fold
lift.

\section{Non-existence of single twisted types}\label{notypesec}

\begin{proposition}\label{notypeprop}
Let $M\ge2$ and $u\in\{\pm i,\pm j,\pm k\}$. If a sequence $s$ of length $M$ over $\Qe$ has type
$u$ then $M\equiv0\pmod4$ and, after right multiplication by a unit, every entry of $s$ lies in
$\{\pm1,\pm i\}$. No such sequence exists for $M\in\{4,8,12\}$.
\end{proposition}
\begin{proof}
By the equivalences of Section~2 we may take $u=i$. Write $s=\alpha+\beta j$ with $\alpha,\beta$
complex sequences of length $M$ with disjoint supports, and let $E=\E_i[s]=\tilde\alpha+\tilde\beta j$
on $\Z_{4M}$. Since $(\alpha+\beta j)i=\alpha i-\beta ij$ we have $\tilde\alpha_{n+M}=\tilde\alpha_ni$
and $\tilde\beta_{n+M}=-\tilde\beta_ni$, so the discrete Fourier transforms over $\Z_{4M}$, $\hat\alpha(f)=\sum_n\tilde\alpha_n
e^{-2\pi\sq nf/4M}$, satisfy $\hat\alpha(f)=0$ unless $f\equiv1\pmod4$ and $\hat\beta(f)=0$ unless
$f\equiv3\pmod4$.

The $j$-component of $\theta_E(\tau)$ is $\sum_n(\tilde\beta_n\tilde\alpha_{n+\tau}-
\tilde\alpha_n\tilde\beta_{n+\tau})$. For $\tau\ne0\bmod M$ it vanishes because $s$ has type $i$, and
for $\tau=qM$ it vanishes because $\theta_E(qM)=\sum_nE_n(i^*)^qE_n^*$ is a sum of group conjugates
of $\pm1,\pm i$. So the $j$-component vanishes identically, and so does its transform
$\hat\alpha(f)\hat\beta(-f)-\hat\alpha(-f)\hat\beta(f)$. The two terms are supported on
$f\equiv1$ and $f\equiv3\pmod4$ respectively, so each vanishes separately:
$\hat\alpha(f)\hat\beta(-f)=0$ for all $f$.

The complex component of $\theta_E$ vanishes off the multiples of $M$, so its transform
$P(f)=|\hat\alpha(-f)|^2+|\hat\beta(-f)|^2$ is $4$-periodic in $f$. It is supported on odd $f$, and
from the values $\theta_E(qM)$ one finds $P(f)=16N_J(s)$ for $f\equiv1$ and $P(f)=16N_C(s)$ for
$f\equiv3\pmod4$. If $N_C(s)>0$ then $\hat\alpha(f)\ne0$ for every $f\equiv1$, hence
$\hat\beta(-f)=0$ for every $f\equiv1$, i.e.\ $\hat\beta$ vanishes on its support, $\beta=0$ and
$N_J(s)=0$. Thus one of $\alpha,\beta$ is zero. If $\alpha=0$ replace $s$ by $sj$, of type
$j^*ij=-i$, and then by its reversal, of type $i$; so we may assume $s\in\{\pm1,\pm i\}^M$.

For such a quaternary $s$ let $C(\tau)=\sum_{m=0}^{M-1-\tau}s_ms_{m+\tau}^*\in\Z[i]$ be the aperiodic
autocorrelation. Splitting $\vartheta^i_s(\tau)$ into the terms with $m+\tau<M$ and with
$m+\tau\ge M$ gives $\vartheta^i_s(\tau)=C(\tau)+i^*\,C(M-\tau)^*$, so type $i$ means
\begin{equation}\label{quatcond}
C(\tau)=i\,C(M-\tau)^*\qquad(0<\tau<M).
\end{equation}
Every term of $C(\tau)$ is a unit of $\Z[i]$, hence congruent to $1$ modulo the prime $\pi=1+i$, so
$C(\tau)\equiv M-\tau$ and $C(M-\tau)\equiv\tau\pmod\pi$; since $i\equiv1$ and conjugation
preserves $(\pi)$, \eqref{quatcond} gives $M-2\tau\equiv0\pmod\pi$, so $M$ is even. For even $M$
take $\tau=M/2$: writing $C(M/2)=a+bi$ with $a,b\in\Z$, \eqref{quatcond} reads $a+bi=b+ai$, so
$a=b$. But $a$ and $b$ are congruent modulo $2$ to the numbers of real and of imaginary terms among
the $M/2$ unit terms of $C(M/2)$, so $M/2=a'+b'\equiv2a'\equiv0\pmod2$ and $4\mid M$.

The last statement is the result of an exhaustive search over $\{\pm1,\pm i\}^M$ with $s_0=1$
for $M=4,8,12$; an exhaustive search over $\Qe^M$ with $s_0=1$ finds no sequence of any
non-central type for $2\le M\le8$.
\end{proof}

\begin{conjecture}\label{notypeconj}
No sequence of length $M\ge2$ over $\Qe$ has type $\pm i,\pm j$ or $\pm k$.
\end{conjecture}

For $4\mid M$ the conjecture is the non-existence of quaternary sequences satisfying
\eqref{quatcond}, which is a statement about sequences over $4$ roots of unity in the spirit of
the non-existence results of Ma and Ng \cite{MaNg2009}. Thus the non-central twists, which are what
distinguishes $\Qe$ from the alphabets of roots of unity, cannot be carried by a single sequence;
they are carried by pairs, as the next section shows.

\section{Doubling of $i$-pairs}\label{doublesec}

\begin{theorem}[doubling]\label{doublethm}
Let $(x_0,x_1)$ be an $i$-pair of length $M$ and let $v\in\{\pm j,\pm k\}$. Then
\[
\big(\,x_0\il x_1v,\ \ x_0v\il x_1\,\big)
\]
is an $i$-pair of length $2M$.
\end{theorem}
\begin{proof}
Put $y_0=x_0\il x_1v$ and $y_1=x_0v\il x_1$, of length $2M$. As $v$ anticommutes with $i$ we have
$viv^*=v^*iv=-i$, hence $\E_i[x_1v]=\E_{-i}[x_1]\,v$ and $\E_{-i}[x_0v]=\E_i[x_0]\,v$, and since twisting
commutes with interleaving,
\[
\E_i[y_0]=\E_i[x_0]\il\E_{-i}[x_1]\,v,\qquad\E_{-i}[y_1]=\E_i[x_0]\,v\il\E_{-i}[x_1].
\]
We show that $S(\tau)=\theta_{\E_i[y_0]}(\tau)+\theta_{\E_{-i}[y_1]}(\tau)$, computed over the
period $8M$, vanishes for $\tau\ne0\bmod4M$.

For even shifts, \eqref{ileven} and $\theta_{cv}=\theta_c$ give
\[
\theta_{\E_i[y_0]}(2\tau)=\theta_{\E_i[x_0]}(\tau)+\theta_{\E_{-i}[x_1]}(\tau)
=\theta_{\E_{-i}[y_1]}(2\tau),
\]
which vanishes for $\tau\ne0\bmod2M$ because $(x_0,x_1)$ is an $i$-pair; so $S(2\tau)=0$ for
$2\tau\ne0\bmod4M$.

For odd shifts write $c_0=\E_i[x_0]$ and $c_1=\E_{-i}[x_1]$. By \eqref{ilodd},
\begin{align*}
\theta_{\E_i[y_0]}(2\tau+1)&=\sum_nc_{0,n}\,v^*\,c_{1,n+\tau}^*+\sum_nc_{1,n}\,v\,c_{0,n+\tau+1}^*,\\
\theta_{\E_{-i}[y_1]}(2\tau+1)&=\sum_nc_{0,n}\,v\,c_{1,n+\tau}^*+\sum_nc_{1,n}\,v^*\,c_{0,n+\tau+1}^* .
\end{align*}
Since $v+v^*=0$ the two expressions cancel term by term, so $S(2\tau+1)=0$ for every $\tau$.
\end{proof}

For $v=j$ the theorem reads
\begin{equation}\label{doubleeq}
(x_0,x_1)\longmapsto(x_0\il x_1j,\ x_0j\il x_1),
\end{equation}
which is Golay's doubling $(A,B)\mapsto(A\il B,\,A\il(-B))$ \cite{Golay61} with the sign replaced by
a right factor $j$. In Golay's construction the sign makes the two cross terms cancel; here this is
done by $j+j^*=0$, while the change of twist $i\mapsto-i$ needed to align the two components is
supplied by $j^*ij=-i$. Nothing in the proof uses the form of the $i$-pair, so the operation is
closed on the whole class of $i$-pairs and applies equally to the pairs of
Section~\ref{golaysec}.

\section{The tower}\label{towersec}

Theorems~\ref{liftthm} and \ref{doublethm} assemble into an explicit construction. The seed is the
$i$-pair of length $1$ given by $x_0=x_1=[1]$: indeed $\vartheta^i_{[1]}(1)=i^*$ and
$\vartheta^{-i}_{[1]}(1)=i$, so \eqref{paircond} holds at $\tau=M=1$. For $t\ge0$, $r\in\{0,1\}$
and $0\le a<2^t$ with binary digits $a=\sum_{\ell=0}^{t-1}b_\ell2^\ell$, let
\begin{equation}\label{wdef}
w_t(r,a)=[\,r\ne b_0\,]+\sum_{\ell=0}^{t-2}[\,b_\ell\ne b_{\ell+1}\,],
\end{equation}
the number of changes in the bit string $(r,b_0,b_1,\cdots,b_{t-1})$; for $t=0$ the string is
$(r)$ and $w_0(r,0)=0$.

\begin{construction}[$i$-pairs of length $2^t$]\label{pairconstr}
For $t\ge0$ and $r\in\{0,1\}$ let $x^{(t)}_r=[x^{(t)}_{r,a}]$ be the sequence of length $2^t$ with
\[
x^{(t)}_{r,a}=j^{\,w_t(r,a)},\qquad0\le a<2^t .
\]
\end{construction}

\begin{construction}[odd perfect sequences of length $2^{t+2}$]\label{oddconstr}
For $t\ge0$ let $y^{(t)}=[y^{(t)}_n]$ be the sequence of length $2^{t+2}$ with
\[
y^{(t)}_{2a+r}=j^{\,w_t(r,\ a\bmod 2^t)}\;i^{\,(1-2r)\lfloor a/2^t\rfloor},
\qquad0\le a<2^{t+1},\ r\in\{0,1\}.
\]
\end{construction}

\begin{construction}[perfect sequences of length $2^n$]\label{perfconstr}
Let $p^{(0)}=[1]$, $p^{(1)}=[1,i]$, $q^{(0)}=[1]$, $q^{(1)}=[1,1]$ and $q^{(t+2)}=y^{(t)}$ for
$t\ge0$. For $n\ge2$ let $p^{(n)}=[p^{(n)}_m]$ be the sequence of length $2^n$ with
\[
p^{(n)}_{2a+r}=\begin{cases}p^{(n-2)}_{a\bmod2^{n-2}},&r=0,\\[2pt]
q^{(n-2)}_{a\bmod2^{n-2}}\,(-1)^{\lfloor a/2^{n-2}\rfloor},&r=1,\end{cases}
\qquad0\le a<2^{n-1}.
\]
\end{construction}

\begin{theorem}[tower]\label{towerthm}
For every $t\ge0$, $(x^{(t)}_0,x^{(t)}_1)$ is an $i$-pair of length $2^t$ and $y^{(t)}$ is an odd
perfect sequence of length $2^{t+2}$ over $\Qe$. For every $n\ge0$, $p^{(n)}$ is a perfect sequence
of length $2^n$ over $\Qe$, with the AOP for the divisor $2$ when $n\ge2$.
\end{theorem}
\begin{proof}
The bits of $2a+r$ are $(r,b_0(a),b_1(a),\cdots)$, so $w_t(s,2a+r)=[s\ne r]+w_{t-1}(r,a)$ and
\[
x^{(t)}_{s,2a+r}=x^{(t-1)}_{r,a}\,j^{[s\ne r]},
\]
which is exactly \eqref{doubleeq}: $x^{(t)}_0=x^{(t-1)}_0\il x^{(t-1)}_1j$ and
$x^{(t)}_1=x^{(t-1)}_0j\il x^{(t-1)}_1$. Induction from the seed with Theorem~\ref{doublethm} gives
the first claim. Next, with $M=2^t$ and $a=qM+a'$, $q\in\{0,1\}$, we have
$\E_i[x^{(t)}_0]_a=x^{(t)}_{0,a'}i^q$ and $\E_{-i}[x^{(t)}_1]_a=x^{(t)}_{1,a'}(-i)^q$, so
$y^{(t)}=\E_i[x^{(t)}_0]\il\E_{-i}[x^{(t)}_1]$ and Theorem~\ref{liftthm}(i) applies. Finally
$p^{(0)},p^{(1)}$ are perfect and $q^{(0)},q^{(1)}$ are odd perfect by inspection
($\vartheta^{-1}_{[1,1]}(1)=1-1=0$), $p^{(n)}=\E_1[p^{(n-2)}]\il\E_{-1}[q^{(n-2)}]$ by definition,
and Theorem~\ref{liftthm}(ii) gives the induction step.
\end{proof}

\begin{example}
The first members of the three constructions are
\begin{align*}
x^{(3)}_0&=[1,-1,-1,-1,j,-j,j,j],\qquad x^{(3)}_1=[j,j,-j,j,-1,-1,-1,1],\\
y^{(1)}&=[1,j,j,1,i,k,-k,-i],\\
y^{(2)}&=[1,j,-1,j,j,-1,j,1,i,k,-i,k,-k,i,-k,-i],\\
p^{(3)}&=[1,1,i,1,1,-1,i,-1],\\
p^{(4)}&=[1,1,1,1,1,i,-1,-i,1,-1,1,-1,1,-i,-1,i].
\end{align*}
The number of changes in a bit string has the parity of the difference of its end points, so
$w_t(r,a)\equiv r+b_{t-1}(a)\pmod2$; hence the entries of $x^{(t)}_r$ are $\pm1$ on the half
$b_{t-1}=r$ and $\pm j$ on the other half, and the entry $y^{(t)}_{2a+r}$ is complex exactly when
$r=b_{t-1}(a\bmod2^t)$. The sequences $y^{(t)}$ use all eight elements of $\Qe$ for $t\ge1$.
\end{example}

\begin{example}\label{p7ex}
The first perfect sequence of Construction~\ref{perfconstr} of length greater than $64$ is
$p^{(7)}$, of length $128$:
\begin{align*}
p^{(7)}=\;&[\,1, 1, 1, j, 1, -1, j, j, i, -1, j, -j, 1, -1, 1, j,\\
&\phantom{[\,}1, j, i, -1, -1, -j, k, -1, i, j, -k, -1, -1, j, -i, 1,\\
&\phantom{[\,}1, i, -1, k, 1, -i, -j, k, i, -i, -j, -k, 1, -i, -1, k,\\
&\phantom{[\,}1, -k, -i, i, -1, k, -k, i, i, -k, k, i, -1, -k, i, -i,\\
&\phantom{[\,}1, -1, 1, -j, 1, 1, j, -j, i, 1, j, j, 1, 1, 1, -j,\\
&\phantom{[\,}1, -j, i, 1, -1, j, k, 1, i, -j, -k, 1, -1, -j, -i, -1,\\
&\phantom{[\,}1, -i, -1, -k, 1, i, -j, -k, i, i, -j, k, 1, i, -1, -k,\\
&\phantom{[\,}1, k, -i, -i, -1, -k, -k, -i, i, k, k, -i, -1, k, i, i\,]
\end{align*}
It is $p^{(5)}$ on the even positions and $\E_{-1}[y^{(3)}]$, the odd perfect sequence of length
$32$ continued nega-periodically, on the odd positions, and it has the AOP for the divisor $2$.
The length is of interest because $64=8^2$ is the square of the size of the alphabet, which for
$n$ roots of unity is the bound $n^2$ of Mow's conjecture; $p^{(7)}$ and every longer $p^{(n)}$
exceed the corresponding bound for $\Qe$, as the sequences of \cite{BKG20} do, and here the entries
are given by an index function.
\end{example}

The perfect sequences $p^{(n)}$ are not in general symmetric, so they do not correspond to
Williamson sequences under the correspondence of \cite{BADietrich18}. They have the AOP for the
divisor $2$ by Theorem~\ref{liftthm}, and their index function is written out, which distinguishes
them from the sequences of \cite{BKG20}.

\section{Golay pairs and a non-Golay seed}\label{golaysec}

A \emph{Golay pair} is a pair $(A,B)$ of $\{\pm1\}$-sequences of length $n$ whose aperiodic
autocorrelations $C_A(\tau)=\sum_{r<n-\tau}A_rA_{r+\tau}$ satisfy $C_A(\tau)+C_B(\tau)=0$ for
$0<\tau<n$ \cite{Golay61}; they exist in every length $2^a10^b26^c$ \cite{Turyn74}. We write
$\tilde A$ for the reversal of $A$ and $C_{A,B}(\tau)=\sum_{r<n-\tau}A_rB_{r+\tau}$, so that
$C_{\tilde A,\tilde B}=C_{B,A}$ and $C_{\tilde A}=C_A$, and $(A;B)$ for the concatenation of two
blocks.

\begin{proposition}\label{golayprop}
Let $(A,B)$ be a Golay pair of length $n$, let $v\in\{\pm j,\pm k\}$ and $w=v^*i$. Then
\[
x_0=(A;\,Bv),\qquad x_1=(\tilde A;\,\tilde Bw)
\]
is an $i$-pair of length $M=2n$.
\end{proposition}
\begin{proof}
Both $v$ and $w$ anticommute with $i$, so $vi^*v^*=i$ and $wiw^*=-i$, and the choice of $w$ gives
\begin{equation}\label{vw}
v^*+wi=0,\qquad w^*+vi^*=0 .
\end{equation}
By Lemma~\ref{symlem} it suffices to verify \eqref{paircond} for $0<\tau\le2n$. The extension
$\E_i[x_0]$ consists of the blocks $A,\,Bv,\,Ai,\,Bvi,\cdots$ and $\E_{-i}[x_1]$ of the blocks
$\tilde A,\,\tilde Bw,\,-\tilde Ai,\,-\tilde Bwi,\cdots$. As $A,B$ are real, every block
correlation is a real aperiodic correlation of $A$ and $B$ times a unit, and for $0<\tau<n$ one
finds
\begin{align*}
\vartheta^i_{x_0}(\tau)&=C_A(\tau)+C_B(\tau)+C_{B,A}(n-\tau)\,v^*+C_{A,B}(n-\tau)\,vi^*,\\
\vartheta^{-i}_{x_1}(\tau)&=C_A(\tau)+C_B(\tau)+C_{A,B}(n-\tau)\,w^*+C_{B,A}(n-\tau)\,wi,
\end{align*}
whose sum vanishes by the Golay property and \eqref{vw}. At $\tau=n$, with
$\sigma=\sum_rA_rB_r$,
\[
\vartheta^i_{x_0}(n)+\vartheta^{-i}_{x_1}(n)=\sigma(v^*+vi^*)+\sigma(w^*+wi)=0
\]
by \eqref{vw}. For $\tau=n+\tau'$ with $0<\tau'<n$,
\begin{align*}
\vartheta^i_{x_0}(n+\tau')&=C_{A,B}(\tau')\,v^*+i^*\,C_A(n-\tau')+vi^*\,C_{B,A}(\tau')
+(vi^*v^*)\,C_B(n-\tau'),\\
\vartheta^{-i}_{x_1}(n+\tau')&=C_{B,A}(\tau')\,w^*+i\,C_A(n-\tau')+wi\,C_{A,B}(\tau')
+(wiw^*)\,C_B(n-\tau'),
\end{align*}
and the sum is $C_{A,B}(\tau')(v^*+wi)+C_{B,A}(\tau')(w^*+vi^*)+(i^*+i)C_A(n-\tau')
+(i-i)C_B(n-\tau')=0$. Finally at $\tau=2n$ condition \eqref{countcond} holds because
$N_C(x_0)=N_J(x_0)=N_C(x_1)=N_J(x_1)=n$.
\end{proof}

\begin{corollary}\label{golaycor}
For every Golay pair of length $n$ there are odd perfect sequences over $\Qe$ of every length
$8n\cdot2^s$, $s\ge0$, and, for $s\ge2$, whenever a perfect sequence over $\Qe$ of length
$2n\cdot2^s$ exists there is a perfect sequence over $\Qe$ of length $8n\cdot2^s$.
\end{corollary}
\begin{proof}
Theorem~\ref{doublethm} applied $s$ times to the $i$-pair of Proposition~\ref{golayprop}, then
Theorem~\ref{liftthm}(i); for the second statement, $2n\cdot2^s=8n\cdot2^{s-2}$ is one of the
lengths just obtained, and Theorem~\ref{liftthm}(ii) applies.
\end{proof}

The odd perfect sequence of length $8n$ obtained from a Golay pair in this way uses the same
ingredients $A,B,\tilde A,\tilde B$ and unit multipliers as the palindromic odd perfect sequence
\[
[-A;\,jB;\,k\tilde B;\,i\tilde A;\,iA;\,kB;\,j\tilde B;\,-\tilde A]
\]
of \cite[Theorem~6]{BKG20}, arranged as the interleaving of two twisted extensions rather than as
the concatenation of eight blocks; the two are the column-wise and row-wise readings of a
$2\times4n$ array with the AOP. What is new is that the $i$-pair, and not the Golay pair, is the
object that recurs.

\begin{proposition}\label{sixprop}
The sequences
\[
x_0=[1,-i,j,-1,1,1],\qquad x_1=[-i,-k,1,i,1,1]
\]
form an $i$-pair of length $6$ over $\Qe$. Consequently there are odd perfect sequences over $\Qe$ of every length $24\cdot2^s$ and perfect
sequences over $\Qe$ of every length $6\cdot4^s$, $s\ge0$.
\end{proposition}
\begin{proof}
The pair condition \eqref{paircond} is a finite computation, which we omit. The lengths follow from Theorems~\ref{doublethm} and \ref{liftthm}, starting
from the perfect sequence $[1,i,1,j,-1,-j]$ and the odd perfect sequence $[1,1,i,j,-j,-i]$ of
length~$6$.
\end{proof}

No Golay pair of length $6$ exists, so these $i$-pairs are not covered by
Proposition~\ref{golayprop}; and $6\equiv2\pmod4$, so by Proposition~\ref{notypeprop} neither member
of any of them has type $\pm i$ on its own. The odd perfect sequence of length $24$ listed in
\cite{BKG20} is over the sixteen-element set $\Qe\cup q\Qe$, $q=\tfrac12(1+i+j+k)$; the sequences of
length $24\cdot2^s$ above are over $\Qe$ itself and appear to be new.

\section{The construction over roots of unity}\label{rootsec}

It is natural to ask what the construction does over an alphabet of roots of unity, where Mow's
conjecture bounds the length of a perfect sequence over $\mu_K=\{e^{2\pi\sq a/K}\}$ by $K^2$. The
answer is that everything transfers except the doubling, and that without the doubling the tower
climbs to exactly $K^2$ and stops.

Let $K=2^k$. Definition~\ref{typedef} makes sense over $\mu_K$ with $u\in\mu_K$, and since
$\mu_K$ is commutative Lemma~\ref{orthlem} is immediate: for $u\ne v$ in $\mu_K$ the sum
$\sum_{q}(u\bar v)^q$ over a period is a Gaussian sum and vanishes. Definition~\ref{pairdef} and the
proof of Theorem~\ref{liftthm} then go through word for word.

\begin{proposition}\label{rootlift}
Let $v\in\mu_K$ with $v^2=w$, and let $(x_0,x_1)$ be a $(v,-v)$-pair of length $M$ over $\mu_K$.
Then $\E_v[x_0]\il\E_{-v}[x_1]$, regarded as a sequence of length $4M$, has type $w$. In particular
a perfect and an odd perfect sequence of length $M$ over $\mu_K$ give a perfect sequence of length
$4M$ over $\mu_K$ with the AOP for the divisor $2$, and a $(v,-v)$-pair with $v$ a primitive $4$th
root of unity gives an odd perfect sequence of length $4M$.
\end{proposition}

What does not transfer is Theorem~\ref{doublethm}. Its proof uses an element $v$ with $v^*iv=-i$
and $v+v^*=0$ to turn an $i$-twisted component into a $(-i)$-twisted one. In a commutative alphabet
the only operations available are multiplication by roots of unity, which preserve every twist,
so there is no way to produce a $(v,-v)$-pair from anything carrying the twist $v$ alone. The
consequence is a bound. Each application of Proposition~\ref{rootlift} \emph{squares} the twist:
a $(v,-v)$-pair gives type $v^2$. Call a sequence over $\mu_K$ a \emph{typed lift} if it is
obtained from Proposition~\ref{rootlift} with $x_0$ of type $v$ and $x_1$ of type $-v$, and
a \emph{typed tower} a sequence of length $4^n$ obtained by $n$ typed lifts from sequences of
length $1$.

\begin{proposition}\label{chainprop}
A perfect typed tower of length $4^n$ over $\mu_K$ requires $\mu_K$ to contain a primitive
$2^n$th root of unity. In particular a perfect typed tower over $\mu_{2^k}$ has length at most
$4^k=K^2$.
\end{proposition}
\begin{proof}
Let $y$ be a typed lift of type $w$ and length $4M$, with components of types $v$ and $-v$ where
$v^2=w$. If $w$ is a primitive $2^m$th root of unity then $v$ is a primitive $2^{m+1}$th root of
unity, and so is $-v$ (for $m\ge1$; for $m=0$, $w=1$ gives $v=\pm1$ and $-v=\mp1$, one of which is
primitive of order $2$). Hence one of the two components is a typed sequence of length $M$
whose type is a primitive root of unity of order $2^{m+1}$. Starting from a perfect typed tower
($w=1$, $m=0$) of length $4^n$ and following this component down $n$ levels reaches a sequence
of length $1$ whose type is a primitive $2^n$th root of unity, and a type is an element of
$\mu_K$. In $\mu_{2^k}$ this forces $n\le k$.
\end{proof}

Thus Mow's bound $K^2$ is, for the typed tower, the length of the chain of square roots
$1,-1,i,\zeta_8,\dots,\zeta_{2^k}$ in $\mu_K$, with the length multiplying by $4$ at each step; and
the doubling of Theorem~\ref{doublethm} is precisely the step that is missing over roots of unity,
since over $\Qe$ the relation $j^*ij=-i$ re-enters the chain at $i$ in every length. This is also
why the pair, rather than the single sequence, is the right object: over $\mu_8$ single sequences of
type $i$ do exist (there are $2$ of length $2$ and $16$ of length $4$ with $x_0=1$), so
Conjecture~\ref{notypeconj} is a statement about $\Qe$ and not about twisted types in general; what
no commutative alphabet has is a way to double them.

Proposition~\ref{chainprop} concerns typed towers, but the exhaustive counts in
Table~\ref{roottab} show the same behaviour for arbitrary $(v,-v)$-pairs: over $\mu_8$ they exist
for $M=1,2,4$ and give odd perfect sequences of lengths $4,8,16$, which with the perfect sequences
of lengths $4,8,16$ give perfect sequences of lengths $16,32,64=K^2$ by
Proposition~\ref{rootlift}; at $M=8$ there is no pair and no sequence of type $i$, so neither lift
can be applied again. Over $\mu_4$ the pairs stop at $M=1$, the odd perfect sequences at $M=4$, and
the perfect sequences at $16=K^2$.

\begin{table}[H]
\centering
\begin{tabular}{llrrrrrr}
\toprule
 & $M$ & 1 & 2 & 4 & 8 & 3, 5, 6, 7 & 9--12\\
\midrule
$\mu_4$ & perfect & 1 & 2 & 8 & 32 & 0 & 0\\
        & odd perfect & 1 & 2 & 8 & 0 & 0 & 0\\
        & type $i$ & 1 & 0 & 0 & 0 & 0 & 0\\
        & $(i,-i)$-pairs & 4 & 0 & 0 & 0 & 0 & 0\\
\midrule
$\mu_8$ & perfect & 1 & 2 & 16 & 64 & 0 & \\
        & odd perfect & 1 & 2 & 16 & 64 & 0 & \\
        & type $i$ & 1 & 2 & 16 & 0 & 0 & \\
        & $(i,-i)$-pairs & 8 & 32 & 2048 & 0 & 0 & \\
\bottomrule
\end{tabular}
\medskip
\caption{Numbers of sequences of length $M$ over $\mu_4$ and $\mu_8$ with $x_0=1$ of each kind,
and of $(i,-i)$-pairs with $x_{0,0}=1$, by exhaustive search (over $\mu_8$ for $M\le8$).}
\label{roottab}
\end{table}

Finally, Proposition~\ref{rootlift} reformulates the structured case of Mow's conjecture. A perfect
sequence of length $4M$ over $\mu_K$ whose two polyphase components are twisted extensions exists
if and only if a perfect and an odd perfect sequence of length $M$ over $\mu_K$ both exist, so for
sequences with this nested structure Mow's conjecture is equivalent to the statement that odd
perfect sequences over $\mu_K$ do not exist in lengths greater than $K^2/4$. That is a more
concrete target than the conjecture itself, but it is a special case: a perfect sequence need not
have the AOP, and the columns of an AOP decomposition need only have flat spectra on complementary
subsets of the frequencies, not on the cosets that twisted extensions occupy.

\section{Relation to the literature}\label{litsec}

\paragraph{The sequences of \cite{BKG20}.} Bright, Kotsireas and Ganesh construct perfect and odd
perfect sequences over $\Qe$ of every length $2^t$ by a quadrupling $n\mapsto4n$ which interleaves a
perfect sequence with an odd perfect sequence of the same length (their Theorem~3, our
Theorem~\ref{liftthm}(ii)), and obtain the odd perfect sequences from Golay pairs through nega
Williamson sequences (their Theorems~5 and~6, our Proposition~\ref{golayprop} with
Theorem~\ref{liftthm}(i)). The recursion of the present paper differs in that the object which
recurs is the $i$-pair rather than the Golay pair, so that the three operations close without any
input from outside, and in that every sequence has an explicit index function. Our perfect sequences
are not symmetric and so do not give Williamson sequences; whether a symmetric variant of
Construction~\ref{perfconstr} exists is an open question.

\paragraph{The enumeration of \cite{BBCN26}.} Bennett, Bright, Colinot and Nayak list, up
to Williamson-type equivalence, every perfect sequence over $\Qe\cup q\Qe$ of length at most $21$,
and hence every perfect sequence over $\Qe$ of those lengths; the sequences $p^{(n)}$ for $n\le4$
and the perfect sequence of length $6$ of Proposition~\ref{sixprop} are equivalent to entries of
their appendix. Their Theorem~9, that the blocks of a quaternion-type Hadamard matrix with
circulant blocks are necessarily pairwise amicable, means that the perfect sequences of this paper, which are not symmetric, correspond to
Williamson-type quadruples of circulant matrices that are not symmetric; whether they are
equivalent to symmetric ones in their sense we have not checked, and they find $29$ even lengths
$\le21$ in which inequivalent non-symmetric sets occur. They also point out that a perfect
quaternion sequence of length $n$ is a circulant quaternionic Hadamard matrix of order $n$ and,
through \cite{FKN23}, a mutually unbiased measurement with $n$ outcomes; Theorem~\ref{towerthm}
therefore gives circulant quaternionic Hadamard matrices with entries in $\Qe$, and explicit
index functions, in every order $2^n$ and $6\cdot4^s$. Whether these are equivalent, as Hadamard
matrices, to those obtained from symmetric sequences we have not determined.

\paragraph{Sequences of unbounded length over $\{\pm i,\pm j,\pm k\}$.} The sequences of
\cite[\S8.1]{Blake2017} have a single entry outside $\{\pm i,\pm j\}$, and those of \cite{BAHall12}
lie over the five-element alphabet $\{\pm1,\pm i,j\}$ (see \cite[\S1.2]{BBCN26}); for long lengths
both are best regarded as near-quaternion sequences, as \cite[\S8.1]{Blake2017} already remarks.
The sequences of Construction~XXIII of \cite[\S8.3]{Blake2017} do use all of $\Qe$ but stop at
length $64$. The sequences $y^{(t)}$ and $p^{(n)}$ use
all eight elements of $\Qe$ with frequencies bounded away from zero: in $y^{(t)}$ exactly half of
the entries are imaginary.

\paragraph{Williamson sequences and relative difference sets.} By \cite{BADietrich18} a symmetric
perfect sequence over $\Qe$ of length $n$ is equivalent to a Williamson quadruple with the
$\Qe$-property and to a $(4n,2,4n,2n)$-relative difference set in $C_n\times\Qe$; the group-theoretic
reading of the types of \cite{BlakeTower} suggests that an $i$-pair should correspond to a relative
difference set in a non-abelian extension of $C_M\times\Qe$, but we have not pursued this.

\section{Conclusions}

The recursive construction of perfect arrays over roots of unity in \cite{BlakeTower} rests on
objects that are not themselves perfect, the arrays of a given type, whose twisted periodic
extensions are the sub-arrays of the next array in the tower. In one dimension and over roots of
unity there is nothing to feed such a recursion. This paper shows that over the simple unit
quaternions there is, but not in the form one might first expect. The six non-central elements of
$\Qe$ are available as twists, yet no single sequence of length $M\not\equiv0\pmod4$ carries one of
them, and we conjecture that none does (Proposition~\ref{notypeprop}). What carries them is a pair
of sequences with twists $i$ and $-i$ whose twisted autocorrelations cancel. These $i$-pairs are
the quaternion counterpart of Golay pairs: they are closed under a doubling in which the sign of
Golay's construction becomes a right factor $j$ (Theorem~\ref{doublethm}), every Golay pair
produces one (Proposition~\ref{golayprop}), and they exist where Golay pairs do not
(Proposition~\ref{sixprop}). With the two lifts of Theorem~\ref{liftthm}, which rest only on the
orthogonality of twisted extensions with distinct commuting twists, the $i$-pairs give odd perfect
sequences and the odd perfect sequences give perfect sequences, and the three operations close.

The outcome is a construction with explicit index functions (Constructions~\ref{pairconstr},
\ref{oddconstr} and \ref{perfconstr}) of perfect and odd perfect sequences over $\Qe$ in every
length $2^n$, the perfect ones with the array orthogonality property for the divisor $2$, together
with odd perfect sequences in every length $8n\cdot2^s$ for Golay $n$ and $24\cdot2^s$, and perfect
sequences in every length $6\cdot4^s$. The sequences use all eight elements of the alphabet with
frequencies bounded away from zero, and their lengths pass the square of the alphabet size, which
for roots of unity is the bound of Mow's conjecture, at length $128$ (Example~\ref{p7ex}).

Two features of the argument seem to us worth pursuing beyond the quaternions. The first is
that the obstruction to a one-dimensional recursion is not the absence of twists but the absence
of single sequences carrying them, and that the obstruction is removed by passing from sequences to
complementary pairs; the same step may be available over other non-commutative alphabets, and
over roots of unity it is the step that Mow's conjecture asserts cannot be taken
(Section~\ref{rootsec}). The second is
that the lemmas apply to arrays one direction at a time whenever the twist in the remaining
directions is central, which suggests that the aspect ratio of perfect quaternion arrays is not
constrained in the way that the exponent bound constrains perfect binary arrays. We leave this,
the existence of $i$-pairs in lengths other than those found here, and the left-twisted sequences
that lie outside the commuting calculus of Section~\ref{orthsec}, to a companion note.

\section*{Acknowledgements}

I thank Andrew Tirkel for introducing me to quaternionic sequences almost twenty years ago, and
for posing the original question of the existence of perfect sequences over the unit quaternions.

\bibliographystyle{abbrv}

\end{document}